\documentclass[prd,twocolumn,showpacs]{revtex4}
\usepackage{hyperref,amssymb,amsmath,mathrsfs,bm,graphicx}
\usepackage{enumitem}
\usepackage{color}
\usepackage{amsfonts,amsmath,amssymb}
\usepackage{mathrsfs}
\usepackage{amscd}
\usepackage{color}
\usepackage{graphicx}
\begin{document}

	\title{WHOM ACTUALLY DO  MULTIPOLE MOMENTS BELONG TO?} 
\author{J. L. Hern\'andez-Pastora\footnote{Universidad de Salamanca, https://ror.org/02f40zc51.  e-mail address: jlhp@usal.es. ORCID: orcig.org/0000-0002-3958-6083}}
\affiliation{Departamento de Matem\'atica Aplicada. Facultad de Ciencias  and  \\
	Instituto Universitario de F\'\i sica Fundamental y Matem\'aticas.\\
	Universidad de Salamanca. Spain}

\begin{abstract}
Using  an integral definition  given in \cite{RMMsource} to calculate  the relativistic multipole moments (RMM), and the ensuing  generalized relativistic Gauss theorem, we prove  that  the evaluation of that volume integral in Erez-Rosen coordinates, leads to a  specific link between the RMM and the source of the exterior space--time, provided we have a global static axisymmetric metric in that coordinate system for any Weyl exterior field. This result  allows to establish a relationship between the RMM and certain volume integral expressions involving the material content of the source from its energy-momentum tensor as well as the interior metric. In particular the relativistic quadrupole moment for the Erez-Rosen space-time is obtained.
\end{abstract}

\maketitle

\section{Introduction}

Starting from an interior metric of a known relativistic source, the gravitational field of  that source is unique, and is described by a solution of the vacuum Einstein equations, which matches satisfactorily on the boundary that delimits the compact gravitational object.  By studying such outer metric it is possible to obtain information about the source, and this is something to which many research papers have been devoted. In particular, one technique that has proved very useful is to use the RMM to describe the field or its gravitational effects on test particles in the presence of such fields. Thus, for example, we were able to distinguish sources by studying gyroscopic precession \cite{gyros}, geodesics \cite{geod}, as well as the study of circular orbits and ISCOs \cite{iscos}, \cite{sanabria}, or  gravitational radiation \cite{RG} or collapse processes \cite{collapse}, or even to obtain vacuum solutions with a prescribed set of RMM \cite{MQ}

On the contrary, if one has at the beginning a known vacuum solution and tries to determine the source, one finds that there are an infinite number of possible interior metrics and distributions of different matter matching with that exterior.  What has been attempted in many lines of research is to try to relate the RMM, which were so useful in describing the external gravitational field, to the source \cite{epjc}. In this line of work a recent result was obtained in \cite{weylsources}, \cite{kerrsource} where interior metrics are computed and properly matched for any of the external Weyl metrics (vacuum solutions with static and axial  symmetries) \cite{weyl},\cite{quevedo},\cite{tesis} as well as stationary axial (in particular Kerr). The relevance of the result is enhanced by the fact that the inner line element is constructed in terms of the external metric functions evaluated on the boundary, so that we relate the interior of the source (as well as its momentum energy tensor by means of the Einstein equations) to the gravitational field.

The question arising here is: whom actually do the RMM belong to? Do they belong to the exterior or interior metric?.  The answer to that question is clear: RMM are quantities belonging to the global solution at the whole space-time, in such a way that  every interior solution appropriately matched with a vacuum solution automatically assume those RMM. Arguments and specific  calculations to prove this assert will be done in the next sections. The fact is that the RMM have been definined in the literature \cite{geroch}, \cite{thorne}, \cite{hansen}, \cite{FHP}, by means of the exterior metric (Geroch, Hansen, Fodor-Hoenselaers-Perjes method, Thorne...) as well as the result obtained in \cite{RMMsource}. A relationship between those quantities and the material content of the source  would provide the RMM with interesting physical meaning.

In a recent paper a  relativistic generalized Gauss theorem (RGGT)   \cite{RMMsource} was presented which allows  to calculate the RMM defined by Geroch \cite{geroch}  and Thorne \cite{thorne}, through a specific integral definition  trying to generalize the newtonian scenario for that quantities in terms of the source. Such a calculation  is possible if we know the expression in harmonic coordinates of the metric at infinity, and then the volume integral  defining the RMM can be calculated, using the RGGT, as an integral over the surface at infinity.

The aim of this paper is to use that volume integral, which can be explicitly calculated once we know the interior metric matching the vacuum Weyl family of solutions, to  link the RMM to the source of the exterior field (to the energy--momentum components as well as the interior metric). Doing so we will be able to connect the RMM that characterize the space--time with physical properties of the source by means of   specific integral expressions  inside the boundary of the compact object. Although very different in its presentation, this program is somehow similar to the one presented in \cite{gurle}.

To achieve our goal we shall extensively use a general method to construct global static axially symmetric solutions to Einstein equations deployed in \cite{weylsources}. A very brief rewiew of this method is presented in the next section, all the details may be found in  that reference.

Section 3 is devoted to explicitely calculate the integral definition of RMM both in its volume or surface integral version, providing so a prove of the relativistic  Gauss theorem. The result obtained previously is use in Section 4 to establish a relationship between RMM and the source, and some examples are outlined in Section 5.


\section{The global static and axisymmetric metric in Erez-Rosen coordinates}

We shall write the  global static and axisymmetric line element in Erez-Rosen coordinates \cite{ER} as:
\begin{equation}
ds^2=-e^{2\sigma} dt^2+e^{2\nu} dr^2+e^{2 \eta} r^2 d\theta^2+e^{-2 \mu} r^2\sin^2\theta d\varphi^2,
\label{ERglobal}
\end{equation}
where the metric functions depend on $r$ and $\theta$, and are defined as follows
\begin{equation}
e^{2\sigma}=\left\lbrace
\begin{matrix}
& e^{2\hat a} Z^2 &  \ , \ r\leq r_{\Sigma} \nonumber \\
& e^{2 \psi} & \ , \ r\geq r_{\Sigma} \nonumber
\end{matrix}
\right. \quad , \  e^{2\nu}=\left\lbrace
\begin{matrix}
& \frac{e^{2\hat g-2\hat a}}{A}  &  \ , \ r\leq r_{\Sigma} \nonumber \\
& e^{-2\psi+2 \hat{\gamma}}& \ , \ r\geq r_{\Sigma} \nonumber
\end{matrix}
\right.
\end{equation}

\begin{equation}
e^{2\eta}=\left\lbrace
\begin{matrix}
& e^{2\hat g-2\hat a}  &  \ , \ r\leq r_{\Sigma} \nonumber \\
& e^{-2\hat{\psi}+2 \hat{\gamma}}& \ , \ r\geq r_{\Sigma} \nonumber
\end{matrix}
\right. \quad , \ e^{-2\mu}=\left\lbrace
\begin{matrix}
& e^{-2\hat a} &  \ , \ r\leq r_{\Sigma} \nonumber \\
& e^{-2\hat{\psi}}& \ , \ r\geq r_{\Sigma} \nonumber
\end{matrix}
\right.
\end{equation}
where the boundary surface of the source is defined by a constant value $r_{\Sigma}$ of the radial coordinate, $r=r_{\Sigma}$.  $\displaystyle{A\equiv 1-\frac{2Mr^2}{r_{\Sigma}^3}}$, and  $Z=\frac 32 A( r_{\Sigma})-\frac 12 A$, $M$ being the mass. $\hat{\gamma}\equiv \gamma-\gamma_s$, $\hat{\psi}\equiv \psi-\psi_s$, where $\psi$, $\gamma$ are any metric functions of the Weyl family of vacuum solutions,  $\gamma_s$, $\psi_s$ being the corresponding metric functions of the Schwarzschild solution. $\hat a$, $\hat g$  are functions in the variables $(r,\theta)$ suitable constructed in \cite{weylsources} in order to guarantee a good physical behaviour of the energy-momentum tensor and the matching (Darmois) conditions.

That function $\hat a=a-a_s$ is constructed in such a way that $a(r_{\Sigma})=\psi(r_{\Sigma})$, $a_s(r_{\Sigma})=\psi_s(r_{\Sigma})$, also  $\hat g=g-g_s$ is such that $g(r_{\Sigma})=\gamma(r_{\Sigma})$, $g_s(r_{\Sigma})=\gamma_s(r_{\Sigma})$.

The general solution for the exterior metric function is (Weyl family) \cite{weyl}, \cite{tesis}
\begin{equation}
\psi=\sum_{n=0}^{\infty}(-1)^{n+1} q_n Q_n(x) P_n(y),
\label{erezrosenfamily}
\end{equation}
where $P_n(\cos \theta)$ are Legendre Polynomials, $Q_n(x)$ (with $x\equiv (r-M)/M$) are Legendre functions of second kind and $q_n$ a set of
arbitrary constants.  The relationship between the canonical Weyl coordinates $\lbrace R, \omega=\cos\Theta\rbrace$ and the Erez-Rosen $\lbrace r, y=\cos\theta \rbrace$ system of coordinates is as follows
\begin{eqnarray}
R&=&\sqrt{(r-M)^2-M^2(1-y^2)},\nonumber \\
\omega&=&y\frac{( r -M)}{R}.
\label{weylcoord}
\end{eqnarray}

In addition, to assure a good behaviour of the physical variables at the center of the inner distribution we shall demand (the field equations  have been used):
$\hat a^{\prime}_0={\hat a}_{,\theta 0}={\hat a }_{,\theta  \theta 0}={\hat a}^{\prime}_{,\theta 0}={\hat a }^{\prime}_{,\theta \theta 0}=0,\
\hat g^{\prime}_0={\hat g}_{,\theta 0}={\hat g }_{,\theta \theta 0}={\hat g}^{\prime}_{,\theta 0}={\hat g }^{\prime}_{,\theta \theta 0}=0, 
\hat g^{\prime \prime}_0={\hat g}^{\prime \prime}_{,\theta 0}=0$, 
where prime denotes derivative with respect to  $r$, and the subscript $\theta$ denotes derivative with respect to $\theta$,
and the subscript $0$ indicates that the quantity is evaluated at the center of the distribution.
With all these considerations  we get for the  metric functions the following expressions \cite{weylsources}:
\begin{eqnarray}
\hat a(r,\theta)&=&\hat \psi_{\Sigma} s^2(3-2s)   +r_{\Sigma}\hat \psi^{\prime}_{\Sigma}s^2(s-1)+{\mathbb F},\nonumber \\
\hat g(r,\theta)&=&\hat \Gamma_{\Sigma} s^3(4-3s)   +r_{\Sigma}\hat \Gamma^{\prime}_{\Sigma}s^3(s-1)+{\mathbb G}.
\label{aygsimple}
\end{eqnarray}
with $s\equiv r/r_{\Sigma} \in \left[0,1\right]$ and  ${\mathbb F} \equiv (r-r_{\Sigma})^2F(r,\theta)$, ${\mathbb G}\equiv (r-r_{\Sigma})^2G(r,\theta)$ arbitrary functions with the following constraints: $F(0,\theta)=F^{\prime}(0,\theta)=0$,  $G(0,\theta)=G^{\prime}(0,\theta)=G^{\prime \prime}(0,\theta)=0$.

These metric functions, satisfy the junction conditions   and produce physical variables which are regular within the fluid distribution.  Furthermore  the vanishing of  $\hat g$ on the axis of symmetry, as required by the regularity conditions,  necessary to ensure elementary flatness in the vicinity of  the axis of symmetry, and in particular at the center,
is assured by the fact that $\hat \Gamma_{\Sigma}$ and $\hat \Gamma^{\prime}_{\Sigma}$ vanish on the axis of symmetry.

Even more, at this level of generality we can assure that the junction conditions  imply the vanishing of the radial pressure $(P_{rr}\equiv g_{rr}T^1_1)_{\Sigma}=0$ at the boundary, and it can be shown that  $T_1^2$ vanishes on the boundary surface as well \cite{weylsources}.

When $\hat a=\hat g=0$ we recover the spherical case of a perfect fluid with isotropic pressures:
\begin{eqnarray}
ds^2_I&=& -Z(r)^2 dt^2+ \frac{1}{A( r)} d{ r}^2+{ r}^2 (d \theta^2+\sin^2\theta d\varphi^2), \\
ds^2_E&=& -\left(1-\frac{2M}{r}\right) dt^2+ \frac{ d{ r}^2}{1-\frac{2M}{r}}+{r}^2( d \theta^2+\sin^2\theta d\varphi^2).\nonumber
\label{sphericalER}
\end{eqnarray}

Thus, the global line element (\ref{ERglobal}) describes in the vacuum any solution of the Weyl family ($\psi, \gamma$) and a good behaved  interior solution with an isotropic perfect fluid limit when matching with Schwarzschild space--time.

In \cite{weylsources} the case $F=G=0$ was studied for some examples, in particular the resulting sources for the exterior field of the MQ$^{1}$  \cite{MQ} and Zipoy-Vorhees \cite{zipoy}, \cite{vor} solutions.

Now, the point is that, for any exterior gravitational field an infinite number of sources exist. Accordingly the obvious question arises: how can we relate the possible sources of a given exterior solution belonging to the Weyl family with the multipole structure of the latter? \ In what follows we shall see how to answer to the above question by using the RGGT and the definition of RMM given in \cite{RMMsource}.


\section{The integral definition of RMM}

A definition of multipole moments  was introduced in \cite{RMMsource} for axially symmetric space-times by means of the following integral
\begin{equation}
I_n= \frac{1}{4\pi} \int_V \left[ H_n \hat{\triangle} \xi -\xi \hat{\triangle}(H_n)\right] \sqrt{\hat g}d^3{\vec x},
\label{MMdef}
\end{equation}
where the volume of integration must be extended to the whole space, and the following notation is used:
\begin{equation}
H_n \equiv \frac{(2n-1)!!}{n!} x^{i_1i_2..i_n} e_{i_1i_2..i_n} \qquad , \qquad \xi \equiv\sqrt{-g_{00}},
\end{equation}
where $e^{i_1i_2..i_n}\equiv (e^{i_1}e^{i_2}...e^{i_n})^{TF}$ with  $e^k$ being the unit vector along the positive direction of the symmetry axis, and $TF$ denoting its  trace free  part.  The  Laplacian operator is denoted by $\hat{\triangle}\equiv \frac{1}{\sqrt{\hat g}} \partial_k\left(\sqrt{\hat g} \hat g^{kj}\partial_j \right)$, and $\hat g$ is the determinant of the three-dimensional metric.

Now, the crucial point here is that the integrand in (\ref{MMdef}) is a divergence (see \cite{RMMsource} for details)
\begin{equation}\left[ H_n \hat{\triangle} \xi -\xi \hat{\triangle}(H_n)\right] \sqrt{\hat g}= \partial_k\left[ \sqrt{\hat g} \left(H_n \hat g^{kj} \partial_j \xi-\xi \hat g^{kj} \partial_j H_n\right)\right], \label{divergencia}
\end{equation}
and accordingly, that integral can be evaluated either as a volume integral (\ref{MMdef}), or as a surface  integral:
\begin{equation}
I_n= \frac{1}{4\pi} \int_{\partial V} \left[H_n \hat g^{kj} \partial_j \xi-\xi \hat g^{kj} \partial_j H_n\right]  d\sigma_k,
\label{floworvolu}
\end{equation}
$\partial V$ being the boundary  and $d \sigma_k$ its corresponding surface element.

Two comments are in order at this point:
\begin{itemize}
	\item In the weak field limit  limit $\hat{\triangle}\sim \triangle_f$ (where  $\triangle_f$ denotes the Laplacian in flat space--time) and  $H_n=R^n P_n(\omega)$, and hence the second term in (\ref{MMdef}) vanishes since $\triangle_f\left[R^n P_n(\omega)\right]=0$. Furthermore, in the same weak field limit  $e^{\Psi}\sim 1+\Phi$,  $\Phi$ being the Newtonian potential which verifies the Poisson equation $\triangle_f \Phi=\rho$, and therefore equation (\ref{MMdef}) turns out to be the classical newtonian moments
	\begin{equation}
	I_n=M_n^N= \frac{1}{4\pi} \int_V R^n P_n \rho \  dV.
	\label{NMM}
	\end{equation}
	
	\item As it was shown in \cite{RMMsource} these integrals calculated in harmonic coordinates through the surface integral (\ref{floworvolu}) recover the RMM defined by Geroch (up to a known specific factor).
\end{itemize}

\subsection{Calculation of the surface integral}

We shall now proceed to evaluate the integral expression (\ref{MMdef}) by means of  the  surface integral (\ref{floworvolu}).
First we need to calculate $H_n$, and we obtain:
\begin{eqnarray}
H_n&=& (r-M)^{2k} P_{2k}(y)+\nonumber \\
&+&\sum_{i=0}^{k-1} (r-M)^{2i}\left[-M^2(1-y^2)\right]^{k-i} Q_{2i}^{(k)} (y),
\end{eqnarray}
\begin{equation}
Q_{2i}^{(k)} (y)=\sum_{j=0}^i L_{2k,2j}\left(\begin{matrix} k-j\\
k-j-i
\end{matrix}\right) y^{2j},
\end{equation}
where $n=2k$ (only even order index has been taken since the Weyl solutions used to be considered  posses equatorial symmetry),  and $ L_{2k,2j}$ denotes the coefficient of the Legendre polynomial $P_{2k}(x)$ corresponding to  the monomial $x^{2j}$, i.e.  ${\displaystyle P_{2k}(x)=\sum_{j=0}^k  L_{2k,2j} x^{2j} }.$

Similar expression for the odd index can be obtained.  In fact,  we have that $H_{2n}$ becomes the product $R^{2n} P_{2n}(\omega)$ written in Erez-Rosen coordinates, which according to (\ref{weylcoord}) is ${\displaystyle  H_{2n}= \left((r-M)^2-M^2(1-y^2)\right)^{n} P_{2n}\left(\frac{(r-M)y}{R}\right) }$.

It is easy to see that the surface integral  (\ref{floworvolu}) leads to the following flux  evaluated at the infinity surface ${\displaystyle F_n^{\infty}(\psi^{\prime})}$ since the integration is done over all the space, 
\begin{equation}
F_n(\psi^{\prime})=\frac 12\int_{-1}^{1} r  (r-2M) \left(\psi^{\prime} H_n(r)-H_n^{\prime}(r)\right) dy  \ ,
\end{equation}
which may be reduced to
\begin{equation}
F_n(\psi^{\prime})=\frac 12\int_{-1}^{1} r  (r-2M) \psi^{\prime} H_n(r) dy  \ ,
\label{flujoMn}
\end{equation}
since the following integration in the variable $y$ vanishes for any value of the radial coordinate:
\begin{equation}
\int_{-1}^1 H_n^{\prime}(r) dy =0.
\end{equation}

This integral (\ref{flujoMn}) can be explicitely calculated by taking into account that the exterior metric function $\psi$ (with equatorial symmetry) looks like (see equation (\ref{erezrosenfamily})) a series in the Erez-Rosen family of solutions \cite{quevedo} and then
${\displaystyle \hat{\psi^{\prime}}=-\frac 1M \sum_{k>0}^{\infty}q_{2k}\partial_x Q_{2k}(x)P_{2k}(y)}$, and the following relations hold (where ${\displaystyle x= \frac rm -1}$ is used):
\begin{equation}
\int_{-1}^1 H_{2n} P_{2k}(y) dy=\left\lbrace
\begin{matrix}
&2 N_{2n,2k} P_{2k}(x)  & \ , \ k\leq n \nonumber \\
& 0 & \ , \ k>n ,\nonumber
\end{matrix}
\right.
\end{equation}
with ${\displaystyle N_{2n,2k}=\frac{(2n)! M^{2n}}{((2n+2k+1)!!(2n-2k)!!}}$.

Hence, equation (\ref{flujoMn}) evaluated at infinity ${\displaystyle F_n^{\infty}(\psi^{\prime})}$ becomes 
\begin{eqnarray}
&&F_n^{\infty}(\psi^{\prime})=-a_n^s+\nonumber \\
&-&\frac{1}{M} \left[r(r-2M)\sum_{k>1}^{n}q_{2k}N_{2n,2k}\partial_x Q_{2k}(x)P_{2k}(x)\right]_{r_{\infty}}.
\label{flujoinfity}
\end{eqnarray}
The limit at  the radial infinity of the above equation for such term in the index $k$ leads to a factor ${\displaystyle -\frac{2k+1}{4k+1}}$, and so we finally obtain
\begin{equation}
F_n^{\infty}(\psi^{\prime})=-a_n^s+\sum_{k>1}^{n}q_{2k}N_{2k}(2n)\frac{2k+1}{4k+1}.
\end{equation}

Hence we conclude that the definition of RMM (\ref{MMdef}) is coordinate dependent since the use of harmonic coordinates for the calculation of that integral in \cite{RMMsource}  leads to the RMM of Geroch-Hansen, instead of certain combination of Weyl moments ($a_n$ or the corresponding $q_n$ of the Erez-Rosen representation) that we obtain when the calculation of equation (\ref{MMdef}) is performed in Erez-Rosen coordinates. 

Nevertheless, this result allows us to relate the RMM with volume integrals over the source involving the matter distribution and the interior metric as well, by means of the Gauss theorem and the knowledge of the coeficients $a_n$ and $q_n$ in terms of the RMM \cite{MQ}. We address this aim to the section $4$, and now calculate for completeness the volume integral (\ref{MMdef}).

\subsection{Calculation of the volume integral}

In order to show that the Gauss theorem is perfectly satisfied, we proceed now to evaluate the volume integral (\ref{MMdef}) with the global metric (\ref{ERglobal}).

The integral (\ref{MMdef}) for the  volume extended from the boundary to the infinity, $I_n^E$, is
\begin{eqnarray}
&&I_n^E=\int_{r_{\Sigma}}^{\infty}  \frac{r^2}{2} dr \int_{-1}^1 dy \left\lbrace
\frac{\psi_{,\theta} H_{n,\theta}}{r^2}-\frac{H_{n,\theta}}{r^2} \frac{\cos \theta}{\sin\theta}+\nonumber\right. \\
&-&\left.\frac{H_{n,\theta \theta}}{r^2}-\left(\frac{r-2M}{r}\right)\left[\left(-\psi^{\prime}+\frac{2(r-M)}{r(r-2M)}\right)H_n^{\prime}+H_n^{\prime \prime}\right]\nonumber \right.\\
&+&\left.H_n\left[\left(\psi^{\prime \prime} +\frac{2(r-M)\psi^{\prime}}{r(r-2M)}\right)\frac{r-2M}{r}+\frac{\psi_{,\theta \theta}}{r^2}+\frac{\psi_{,\theta}}{r^2}\frac{\cos \theta}{\sin\theta} \right] \right\rbrace \nonumber \\
\label{IE}
\end{eqnarray}
whereas for the interior volume, $I_n^I$, we have
\begin{eqnarray}
&&I_n^I=\frac 12 \int_{0}^{r_{\Sigma}}  r^2 dr \int_{-1}^1 dy \left\lbrace H_n\left[\frac{3M}{r_{\Sigma}^3}-\frac{A^{\prime} \hat a^{\prime}}{4}+\nonumber \right.\right.\\
&+&\left.\left.Z \left(\frac{2\hat a^{\prime}(1-\frac{3Mr^2}{r_{\Sigma}^3})}{r\sqrt A}+\hat a^{\prime \prime} \sqrt A \frac{{\hat a}_{,\theta \theta}+{\hat a}_{,\theta}\frac{\cos\theta}{\sin\theta}}{r^2 \sqrt A}\right) \right]+\nonumber \right.\\
&-&\left. Z\left[\left(-\hat a^{\prime} \sqrt A  +\frac{2(1-\frac{3Mr^2}{r_{\Sigma}^3})}{r\sqrt A}\right)H_n^{\prime} +\sqrt A H_n^{\prime \prime}+\nonumber \right.\right. \\
&+&\left.\left.\frac{H_{n,\theta \theta}+H_{n,\theta}\frac{\cos\theta}{\sin\theta}}{r^2\sqrt A}-\frac{{\hat a}_{,\theta}}{r^2}\frac{H_{n,\theta}}{\sqrt A}\right]\right\rbrace,
\end{eqnarray}
where $A$ and $Z$ are the previously defined functions only depending on the radial coordinate   which are involved in the interior  line element (\ref{ERglobal}).

Since $\psi$ is  solution of the vacuum field equations, we may write
\begin{equation}
\triangle \psi\equiv\left[\psi^{\prime \prime} +\frac{2(r-M)\psi^{\prime}}{r(r-2M)}\right]\frac{r-2M}{r}+\frac{\psi_{,\theta \theta}}{r^2}+\frac{\psi_{,\theta}}{r^2}\frac{\cos \theta}{\sin\theta}=0,
\label{lapla}
\end{equation}
also  $H_{n,\theta \theta}+ H_{n,\theta} \frac{\cos\theta}{\sin\theta}=(1-y^2)\partial_{yy}H_n-2y\partial_yH_n=\\=\partial_y\left[ (1-y^2)\partial_y H_n\right]$, and
\begin{equation}
\int_{-1}^1 H_n^{\prime} dy =\int_{-1}^1 H_n^{\prime \prime} dy=0.
\end{equation}
Using the above expressions, the integrals $I_n^I$ and $I_n^E$  may be simplified further as follows
\begin{equation}
I_n^E=\frac 12 \int_{r_{\Sigma}}^{\infty}  r^2 dr \int_{-1}^1 dy \left[\hat \psi_{,\theta}\frac{ H_{n,\theta}}{r^2}+\left(\frac{r-2M}{r}\right)\hat \psi^{\prime}H_n^{\prime}\right],
\label{VE}
\end{equation}
and
\begin{eqnarray}
&&I_n^I=\frac 12 \int_{0}^{r_{\Sigma}}  r^2 dr \int_{-1}^1 dy H_n\left[\frac{3M}{r_{\Sigma}^3}-\frac{A^{\prime} \hat a^{\prime}}{4} \right]+\nonumber\\
&+&\int_{0}^{r_{\Sigma}}  \frac{r^2}{2} dr \frac{Z}{\sqrt A}\int_{-1}^1 dy\left\lbrace H_n\left[\hat a^{\prime \prime} A+\hat a^{\prime}\left(\frac 2r-\frac{6Mr}{r_{\Sigma}^3}\right)\right]+\nonumber \right.\\
&+&\left. \hat a ^{\prime} A H_n^{\prime} \right\rbrace+\nonumber\\
&+&\frac 12 \int_{0}^{r_{\Sigma}}  dr \frac{Z}{\sqrt A}\int_{-1}^1 dy \left\lbrace H_n \partial_y \left[(1-y^2)\partial_{y}\hat a\right]+\nonumber \right.\\
&+&\left.(1-y^2)\partial_y \hat a \partial_y H_n\right\rbrace.
\label{Vint}
\end{eqnarray}

Next, notice that the  last term of equation (\ref{Vint}) vanishes after   integration in the variable $y$, since
\begin{eqnarray}
&&\int_{-1}^1 dy \left[ H_n \partial_y \left[(1-y^2)\partial_{y}\hat a\right]+(1-y^2)\partial_y \hat a \partial_y H_n \right]=\nonumber \\
&=&H_n (1-y^2)\partial_y \hat a \ \rvert^1_{-1}=0.
\end{eqnarray}
The second term of equation  (\ref{Vint}) can be integrated with respect to the radial coordinate. The integration for those terms with a factor $H_n$ produces
\begin{eqnarray}
&&\int_{0}^{r_{\Sigma}}  dr \left( \hat a^{\prime} C_1+\hat a^{\prime \prime} C_2\right)=\left(\hat a^{\prime} C_2 \ \right)\rvert^{r_{\Sigma}}_{0}-\int_{0}^{r_{\Sigma}}  \hat a^{\prime} \ \alpha \ dr = \nonumber \\
&=&\hat{\psi}^{\prime}_{\Sigma} C_2(r_{\Sigma})-\int_{0}^{r_{\Sigma}}  \hat a^{\prime} \ \alpha \ dr,
\end{eqnarray}
where $C_1\equiv\frac{Z}{\sqrt A}r^2H_n \left(\frac 2r-\frac{6Mr}{r_{\Sigma}^3}\right)$, $\ \alpha\equiv -C_1+C_2^{\prime}$ and $C_2\equiv\frac{Z}{\sqrt A}r^2H_n \left(1-\frac{2Mr^2}{r_{\Sigma}^3}\right)$

In the above   the matching conditions $\hat a(r_{\Sigma})=\hat \psi(r_{\Sigma})$, $\hat a^{\prime}(r_{\Sigma})=\hat \psi^{\prime}(r_{\Sigma})$, as well as the assumed behaviour at the center $\hat a_0=\hat a^{\prime}_0=0$, have been taken into account.
Also, the  evaluation of $\alpha$ produces  $\alpha=H_n M \frac{r^3}{r_{\Sigma}^3}+H_n^{\prime} \frac{Z}{\sqrt A}r^2A$.

Using all these expresions,  the integral for the interior volume finally reduces to
\begin{equation}
I_n^I=\frac 12 \int_{0}^{r_{\Sigma}}  r^2 dr \int_{-1}^1 dy H_n\left[\frac{3M}{r_{\Sigma}^3} \right]+\frac 12\int_{-1}^1 dy \  \hat{\psi}^{\prime}_ {\Sigma} C_2(r_{\Sigma}).
\label{Vint2}
\end{equation}
By comparing equation (\ref{Vint2}) with the flux (\ref{flujoMn}) evaluated at the boundary surface ${\displaystyle F_n^{\Sigma}(\psi^{\prime})} $ we conclude that they are  equal since $C_2(r_{\Sigma})=r_{\Sigma}  (r_{\Sigma}-2M)  H_n(r_{\Sigma})$ and the first term in equation (\ref{Vint2}) is
\begin{equation}
\frac 12 \int_{0}^{r_{\Sigma}}  r^2 dr \int_{-1}^1 dy  H_n \left[\frac{3M}{r_{\Sigma}^3}\right]=\frac{M^{n+1}}{n+1}=-a_n^s,
\end{equation}
where $a_n^s$ are the Weyl coefficients of Schwarzschild, and $\psi^{\prime}_{\Sigma}=\psi^{\prime s}_{\Sigma}+\hat\psi^{\prime}_{\Sigma}=\frac{M}{r_{\Sigma}  (r_{\Sigma}-2M) }+\hat\psi^{\prime}_{\Sigma}$ and hence the equation (\ref{flujoMn}) evaluated at the boundary surface is equivalent to
\begin{eqnarray}
F_n^{\Sigma}(\psi^{\prime})&=&\frac 12\int_{-1}^{1} r_{\Sigma}  (r_{\Sigma}-2M) \psi_{\Sigma}^{\prime} H_n(r_ {\Sigma}) dy=\nonumber \\
&=&-a_n^s+\frac 12\int_{-1}^{1} r_{\Sigma}  (r_{\Sigma}-2M) \hat \psi_{\Sigma}^{\prime} H_n(r_ {\Sigma}) dy,\nonumber\\
\label{flujorE}
\end{eqnarray}
that is to say, the integral extended to the interior volume $I_n^I$ recovers the flux through the boundary $F_n^{\Sigma}(\psi^{\prime})$.
Thus we have verified (as expected)  that  the volume integral extended to  the interior volume delimited by  the boundary  equals the flux integral  through that surface.

Next,  let us calculate the volume integral at the exterior of the source $I_n^E$. To do so, the second term of equation (\ref{VE}) can be integrated in the radial variable leading to
\begin{eqnarray}
&&\frac 12 \int_{-1}^1 [B H_n]_{r_{\infty}} dy -\frac 12\int_{-1}^1 r^2_{\Sigma}A_{\Sigma} \hat\psi^{\prime}_{\Sigma} H_n({\Sigma}) dy+\nonumber \\
&-&\frac 12 \int_{r_{\Sigma}}^{\infty} dr B^{\prime} \int_{-1}^1 H_n dy,
\label{VEprima}
\end{eqnarray}
with $B\equiv r(r-2M)\hat{\psi}^{\prime}$,  and $[B H_n]_{r_{\infty}}$ denoting the value of that function over the surface at infinity, whereas the first term of equation  (\ref{VE}) can be integrated in the angular variable producing
\begin{equation}
-\frac 12 \int_{r_{\Sigma}}^{\infty} dr \int_{-1}^1 H_n \partial_y[(1-y^2)\hat{\psi_y}]dy,
\label{VEpunto}
\end{equation}
Hence, the sum of equations (\ref{VEprima}) and (\ref{VEpunto}) leads to the following expression for the exterior volume integration
\begin{eqnarray}
I_n^E&=&\frac 12 \int_{-1}^1 [B H_n]_{r_{\infty}} dy -\frac 12\int_{-1}^1 r^2_{\Sigma}A_{\Sigma} \hat\psi^{\prime}_{\Sigma} H_n({\Sigma}) dy \equiv \nonumber\\
&\equiv&  F_n^{\infty}(\psi^{\prime})- F_n^{\Sigma}(\psi^{\prime}) ,
\label{sumadepartes}
\end{eqnarray}
since $\hat\psi$ being a solution of the exterior field equations, verifies $\triangle \hat{\psi}=0$, and the contribution of the Schwarzschild term $\psi^{\prime s}$ in the flux is $a_n^s$ whatever the surface of integration is considered.

Therefore,  the definition of RMM (\ref{MMdef}), evaluated as a volume integral, corresponding to  the sum of the quantities $I_n^I$ and $I_n^E$  leads to the surface integral  ${\displaystyle F_n^{\infty}(\psi^{\prime})}$  at the spatial  infinity, which is the result obtained if one performs the definition (\ref{MMdef}) as a surface integral.

The following remarks are in order at this point:
\begin{itemize}
	\item  This result is true for any interior metric function $\hat a$  (with the only constraints derived from the matching conditions and the well-behaviour at the center). Hence the matching conditions arise as the necessary and suficient condition to hold the equivalence between both kinds of integrals (Gauss theorem).
	
	\item   For any source whose interior metric matches appropriately with  a specific  Weyl solution at the exterior, the RMM are  the same in all the cases, and they are the ones  corresponding to that  Weyl solution.  Whatever the matter distribution of the source be, the RMM are the same ones, since they are determined by the exterior metric to which it is matched.

	\item  It seems to be that the RMM are exclusively  related  to  the gravitational exterior field, but in fact that is a result due to the intrinsic characteristic of the own definition (\ref{MMdef}) and its equivalence between volume or surface integral (\ref{divergencia}), (\ref{floworvolu}). 
	
	\item Therefore, the definition  cannot be used  to constrain the interior solution (the source), since this volume integral (\ref{MMdef}) used to calculate the RMM aparently exclude from the integration the source of the gravitational field. However the flux throuhg the boundary surface contains information both from the source and the gravitational field whose RMM are known. This fact will allows us to link the RMM and the source.
	
	As it is known in Newtonian gravity (NG) we can calculate the NMM of a source from its matter distribution by means of a volume integral (\ref{NMM}) and the gravitational exterior field is characterized by those NMM which are fully determined by the physics of the source. Whereas that identification in NG is forthright is not the case for GR, but nevertheless it is possible to connect the RMM with some volume integrals involving the matter distribution and the interior metric, as we shall see in what follows:
\end{itemize}

\section{The relationship between RMM and the source}

In an attempt to relate the matter distribution of the source with the  RMM, we keep in mind that  the Einstein  equations  connect the metric with the energy-momentum tensor, and we have linked the interior metric functions with the exterior ones as clearly exhibited in (\ref{aygsimple}). Thereferore, we can recalculate the integral definition used  \cite{RMMsource} at the same time that we introduce the matter distribution of the source into those integrals by means of the so-called Tolman density
\begin{equation}
\rho_T\equiv \sqrt{-g_{00}}\left(-T_0^0+T_i^i\right) ,
\end{equation}
since we know \cite{RMMsource}, \cite{weylsources} that ${\displaystyle \hat \triangle  \sqrt{-g_{00}}=4\pi \rho_T}$.

With this consideration,  equation (\ref{MMdef}) becomes
\begin{eqnarray}
I_n&=& \int_V H_n \rho_T \sqrt{\hat g} d^3\vec x-\frac{1}{4\pi}\int_V\xi \partial_k\left(\sqrt{\hat g}\hat g^{kj}\partial_j H_n\right) d^3 \vec x \equiv \nonumber \\
&\equiv& T_n+S_n,
\label{tolmanRMM}
\end{eqnarray}
where ${\displaystyle T_n=\int_V H_n \rho_T \sqrt{\hat g} d^3\vec x}$ denotes the part of the integral involving the Tolman density (content material of the distribution), and ${\displaystyle S_n=-\frac{1}{4\pi}\int_V\xi \partial_k\left(\sqrt{\hat g}\hat g^{kj}\partial_j H_n\right) d^3 \vec x}\quad  $ holds with the other  part of the integral.

This expression (\ref{tolmanRMM}) used to define the RMM generalizes the definition of the Tolman mass \cite{tolman} (Monopole $M_0$),or Komar \cite{komar} moments, since $H_0=1$ and the second term $S_n$ vanishes for that case. The volume integrals in (\ref{tolmanRMM}) must be calculated, as we said above, extended to the whole space, but in the first term of this expression $T_n$ we can limit ourself to the interior of the source since the energy momentum tensor in vacuum vanishes (assuming vanishing electromagnatic field). Equivalently this point is in agreement with the fact that $\int_V  H_n \hat{\triangle} \xi \sqrt{\hat g} d^3 \vec x=0$ at the outside of the source from equations (\ref{IE}) and (\ref{lapla}).

As we prove above (\ref{sumadepartes})
\begin{equation}
I_n^E\equiv T_n^E+S_n^E=S_n^E=F_n^{\infty}(\psi^{\prime})-F_n^{\Sigma}(\psi^{\prime})
\end{equation}
where superscript $E$ denotes exterior volume. Since the evaluation of the integral $I_n$ over the whole space leads to the flux at infinity ${\displaystyle F_n^{\infty}(\psi^{\prime})}$, then 

\begin{equation}
F_n^{\infty}(\psi^{\prime})=I_n^I+F_n^{\infty}(\psi^{\prime})-F_n^{\Sigma}(\psi^{\prime}),
\end{equation} where the superscipt $I$ denotes interior volume. Consequently we obtain the following relation
\begin{equation}
T_n^I=T_n=-S_n^I+F_n^{\Sigma}(\psi^{\prime})
\label{sumadeints}
\end{equation}
which is just the conclusion obtained from equations (\ref{Vint2})-(\ref{flujorE}).

The flux over the boundary surface $F_n^{\Sigma}(\psi^{\prime})$ can be obtained from (\ref{flujoinfity}) as follows
\begin{eqnarray}
&&F_n^{\Sigma}(\psi^{\prime})=-a_n^s+\nonumber \\
&-&\tau (r_{\Sigma}-2M)\left[\sum_{k>1}^{n}q_{2k}N_{2n,2k}\partial_x Q_{2k}(x)P_{2k}(x)\right]_{x=\tau-1}.
\label{fujoinfity}
\end{eqnarray}
${\displaystyle \tau\equiv \frac{r_{\Sigma}}{M}}$ being the compactness factor of the source.
Therefore, the flux provides over the surface $r_{\Sigma}$ information about the RMM  since we know the coeficients $q_{2k}$ in terms of the  Weyl coeficients $a_n$ \cite{MQ}, and  these ones in terms of the RMM \cite{MSA}:
\begin{equation}
q_0=1 \ , \quad 
q_2=\frac {15}{2} \frac{M_2}{M^3} \ , \quad q_4=\frac{45}{4}\frac{M_2}{M^3}+\frac{315}{8}\frac{M_4}{M^5} \ , \cdots
\end{equation}

Hence the equation (\ref{sumadeints}) allows us to write each RMM in terms of two kind of volume integrals one of them $T_n$ involving the matter distribution by means of the Tolman density and the other one $S_n^I$ bringing in the interior metric to the evaluation. The first RMM can be obtained as follows:

\begin{eqnarray}
M_0&=& T_0+S_0^I\nonumber \\
M_2&=&\frac{-1}{\tau(\tau-2) \beta_2(\tau)}\left[-\frac{M^3}{3}+T_2+S_2^I \right] \nonumber \\
M_4&=&\frac{2M^2\left(1+12\frac{\beta_2(\tau)}{\beta_4(\tau)}\right)}{7 \tau(\tau-2) \beta_2(\tau)}\left[-\frac{M^3}{3}+T_2+S_2^I \right]+\nonumber\\
&-&\frac{4}{\tau(\tau-2) \beta_4(\tau)}\left[-\frac{M^5}{5}+T_4+S_4^I \right] \nonumber \\
\label{formulas}
\end{eqnarray}
where the following notation has been used:
\begin{equation}
\beta_n(\tau)\equiv \left[ P_n(x)\partial_x Q_n(x)\right]_{x=\tau-1} \ ,
\end{equation}
$P_n(x)$ and $Q_n(x)$ being the Legendre polynomials and the Legendre polynomials of second kind respectively.

In the classical gravitational analogy the multipole momentum is obtained only as an integral over the source because we do not have in newtonian gravityy  an interior metric. As already seen in \cite{weylsources} the integral $T_0$ gives the mass for any source properly attached to a Weyl exterior, while $S_0^I$ is identically zero since $H_0=1$.

These expressions can be understood in two alternative ways: a qualitative one in which the formulas (\ref{formulas}) should be read in the sense of explaining in what way the source participates in the definition of each RMM, or which characteristics of the source (its density, anisotropic pressure, and the interior metric itself) contribute to the construction of the RMM. We assume that  such moments  are already known from any of the historically defined calculation methods by means of the exterior metric. 

Alternatively, the explicit knowledge of the inner metric allows to calculate the RMM in terms of the physical characteristics of the source.
Evidently the energy-momentum tensor is related through the Einstein equations to the inner metric so that this distinction between the two types of integrals $T_n$ and $S_n^I$ is formal, although significant as we will see in the next section. This quantitative point of view of the formulae (\ref{formulas}) provide us with  an explicit calculation of the RMM using the source itself  rather than the exterior metric.

\section{Some examples}

\subsection{The global model for any Weyl solution}

The equation (\ref{formulas}) is still general for any source appropriately matched to any Weyl exterior solution, since the flux which generates the combinations of RMM was calculated with the vacuum solution. Now we compute the volume integral expresions $T_n$ and $S_n^I$ with a global model metric including all the admissible sources for any axially symmetric vacuum gravitational field in the Weyl gauge \cite{weylsources}. It is easy to see from (\ref{Vint}) that the integrals $S_n^I$ result to be ($V_I$ denotes the volume limited by the boundary surface of the source)
\begin{eqnarray}
&&S_n^I\equiv-\frac{1}{4\pi}\int_{V_I}\xi \partial_k\left(\sqrt{\hat g}\hat g^{kj}\partial_j H_n\right) d^3 \vec x= \nonumber  \\
&=&\frac 12\int_{0}^{r_{\Sigma}}  r^2 dr \int_{-1}^1 dy \frac{Z}{\sqrt{A}}\left[\hat a^{\prime} A H_n^{\prime}+ \frac{(1-y^2)}{r^2} \partial_y \hat a \partial_y H_n \right]\nonumber \\
\label{second}
\end{eqnarray}

With respect to the volume integral $T_n$ involving the Tolman density $\rho_T$ we have that
\begin{eqnarray}
T_n&\equiv& \int_V H_n \rho_T \sqrt{\hat g} d^3\vec x=\nonumber \\
&=&2 \pi \int_{0}^{r_{\Sigma}}  r^2 dr \int_{-1}^1 dy \frac{Z}{\sqrt{A}}H_n\left[(\mu+3P)-\frac{E}{8 \pi} \right]
\label{first}
\end{eqnarray}
where we have used the following notation for the energy momentum tensor
(see \cite{weylsources} for details\footnote{Please take into account  a missprint in that paper for the expression of  $E$ and $\hat p_{zz}$, as well as in the formula (24) in that paper derived from the previous mistaken formulae: the second derivative of the function $\hat g$ with respect to the variable $s$ must contain a forgotten  factor $A$. Same missprints are reproduced in \cite{kerrsource}. The calculations and conclusions derived in both papers are still appropriated and right, since it is a matter of a missprint in the edition of the  latex version.} ):
\begin{eqnarray}
-T^0_0&=&\kappa \left(8 \pi \mu+\hat p_{zz}-E\right),\nonumber\\
T^1_1&=& \kappa \left(8 \pi P-\hat p_{xx}\right),\nonumber \\
T^2_2&=& \kappa \left(8 \pi P+\hat p_{xx}\right),\nonumber \\
T^3_3&=&\kappa \left(8 \pi P-\hat p_{zz}\right),\nonumber \\
T^2_1&=&-\frac{\kappa}{r^2} \hat p_{xy} \qquad , \qquad \kappa\equiv \frac{e^{2\hat a-2\hat g}}{8 \pi}
\label{eegeneral}
\end{eqnarray}
with
\begin{eqnarray}
&&E=-2 \Delta \hat a+(1-A)\left[2 \frac{\hat a^{\prime}}{r}\frac{9 \sqrt{A_{\Sigma}}-4 \sqrt{A}}{3 \sqrt{A_{\Sigma}}- \sqrt{A}}+2 \hat a^{\prime \prime}\right],\nonumber\\
&&\Delta \hat a= \hat a^{\prime \prime}+2\frac{\hat a^{\prime}}{r}+\frac{{\hat a}_{,\theta \theta} }{r^2}+\frac{{\hat a}_{,\theta} }{r^2}\frac{\cos \theta}{\sin \theta},\nonumber \\
&&\hat p_{xx}=-\frac{{\hat a}_{,\theta} ^2}{r^2}-\frac{\hat g^{\prime}}{r}+\hat a^{\prime 2}+\frac{{\hat g}_{,\theta} }{r^2}\frac{\cos \theta}{\sin \theta}+\nonumber \\
&+&(1-A)\left[2 \frac{\hat a^{\prime}}{r}\frac{\sqrt{A}}{3 \sqrt{A_{\Sigma}}- \sqrt{A}}- \hat a^{\prime 2} +\frac{\hat g^{\prime}}{r}\frac{3 \sqrt{A_{\Sigma}}-2 \sqrt{A}}{3 \sqrt{A_{\Sigma}}-\sqrt A}\right], \nonumber \\
&&\hat p_{zz}=-\frac{{\hat a}^2_{,\theta} }{r^2}-\frac{\hat g^{\prime}}{r}-\hat a^{\prime 2}-\frac{{\hat g}_{,\theta \theta} }{r^2}-\hat g^{\prime \prime}+\nonumber\\
&+&(1-A)\left[-2 \frac{\hat a^{\prime}}{r}\frac{\sqrt{A}}{3 \sqrt{A_{\Sigma}}- \sqrt{A}}+ \hat a^{\prime 2} +2\frac{\hat g^{\prime}}{r}+\hat g^{\prime \prime}\right], \nonumber \\
&&\hat p_{xy}=2{\hat a}_{,\theta}\hat a^{\prime}-\hat g^{\prime}\frac{\cos(\theta)}{\sin(\theta)}  -\frac{{\hat g}_{,\theta}}{r} +\frac{(1-A)(2{\hat a}_{,\theta}-{\hat g}_{,\theta})}{r \sqrt A (3 \sqrt{A_{\Sigma}}-\sqrt A)}\nonumber \\
\label{eegeneraldet}
\end{eqnarray}
where $\mu$ and $P$ are the homogeneous and isotropic density and pressure respectively:
\begin{equation}
P=\mu\left(\frac{\sqrt A-\sqrt{A_{\Sigma}}}{3\sqrt{ A_{\Sigma}}-\sqrt A} \right)
\end{equation}
If we rewrite the above expression (\ref{first}) in terms of the physcial parameters of the source we have that 
\begin{equation}
T_n=\frac{M^{n+1}}{n+1}- \frac 14\int_{0}^{r_{\Sigma}}  r^2 dr \int_{-1}^1 dy \frac{\mu}{\mu+3P}H_n E
\label{firstfisical}
\end{equation}
where
\begin{eqnarray}
E&=&-2\left(1-\frac 83 \pi r^2 \mu \right)\hat a^{\prime \prime}- \frac{\hat 4 a^{\prime}}{r} \left(1-\frac 23 \pi r^2 (5\mu-3P)\right)+\nonumber \\
&-&\frac{2}{r^2}\partial_y\left[(1-y^2)\partial_y \hat a\right]
\label{E8pi}
\end{eqnarray}

In \cite{epjc} (see for details therein) the physical characteristic of the source arising from the energy-momentum tensor are expressed in terms of the following pressures and anistropies ${\displaystyle T_m\equiv \frac{T_1^1+T_2^2}{2}, \Pi_{31}, \Pi_{23}, \Pi_{xy}=\frac{-T_m}{8\pi P r^2}\hat p_{xy}}\qquad  $ with the notation ${\displaystyle \Pi_{ij}\equiv T_i^i-T_j^j}$ and the interior metric functions are related as follows ${\displaystyle \hat g=\hat a-\frac 12 \ln\left(\frac{T_m}{P}\right)}$

Hence, the contributions of the integrals $T_n$ to the RMM are due to the inhomogneity of the density along with the pressure $T_m$.

\subsection{Sources of Schwarzschild space-time}

When the particular spherical case is regarded, both the isotropic or anisotropic spherical sources of Schwarzschild vacuum solution lead to the following metric function
\begin{equation}
\hat a=\frac 12 \ln\left(\frac{T_m}{P}\right)
\label{ahat}
\end{equation}
since $\hat g=0$, $\Pi_{xy}=0$, and the unique non vanishing anistropic pressure is ${\displaystyle \Pi_{31}=\frac{T_m}{4 \pi P}\hat p_{xx} }$ since $\hat p_{xx}=-\hat p_{zz}$.

{\bf A) } Now, on the one hand the isotropic perfect fluid subcase  requires $\hat p_{xx}=0$ leading to $T_1^1=T_2^2=T_3^3$
The only
possible solution satisfying at the same time the junction
conditions, is $\hat a=0$ and therefore we get  the
vanishing of the anisotropy $\Pi_{31}=0$ as well as the isotropic
perfect fluid limit  $T_m = P$.

In this particular case (\ref{second}) vanish for any value of the index, i.e. $S_n^I =0$ and (\ref{firstfisical}) $T_n=\frac{M^{n+1}}{n+1}$ and  hence the unique multipole moment is the mass $M_0=M$ from the equations  (\ref{formulas}).

{\bf B) }  Indeed, the same result must be obtained, on the other hand for anistropic sources, since the exterior solution is also Schwarzschild space-time. In fact the spherical case considered $\hat a=\hat a(r)$ and the equation (22) lead to $S_n^I=0$ (\ref{second}) as well. With respect to the integrals $T_n$ (\ref{firstfisical}) the independence of the physical parameters in the angular variable leads to the  following expression
\begin{equation}
T_n=\frac{M^{n+1}}{n+1}-\frac{M^{n}}{2(n+1)}\int_{0}^{r_{\Sigma}}  r^2 dr  \frac{\mu}{\mu+3P} E
\label{constrainprevio}
\end{equation}
and hence the following constraint arises from (\ref{formulas}) since all RMM higher than $M_0$ must vanish:
\begin{eqnarray}
0&=& \int_{0}^{r_{\Sigma}}  r^2 dr  \frac{\mu}{\mu+3P}\left[\left(1-\frac 83 \pi r^2 \mu \right)\hat a^{\prime \prime}+ \nonumber\right.\\
&+&\left.\frac{2 \hat a^{\prime}}{r} \left(1-\frac 23 \pi r^2 (5\mu-3P)\right)\right] 
\label{constrain}
\end{eqnarray}
where the expression for $E$ (\ref{E8pi}) has been considered, and the derivatives of $\hat a$ are obtained from (\ref{ahat}):
\begin{equation}
\hat a^{\prime}=\frac 12 \left( \frac{T_m^{\prime}}{T_m}-\frac{P^{\prime}}{P}\right)\ , \ \hat a^{\prime \prime}=\frac 12\left(\frac{T_m^{\prime\prime}}{T_m}-\frac{T_m^{\prime 2}}{T_m^2}-\frac{P^{\prime\prime}}{P}+\frac{P^{\prime 2}}{P^2}\right)
\end{equation}

Thus the above equation (\ref{constrain}) implies a condition over  the pressure $T_m$ or equivalently the anitropy of the source.

Nevertheless that condition is fullfilled and the equation (\ref{constrain}) is just an identity, since the integral results to be
\begin{eqnarray}
\left[\frac{\mu(\mu+3P)}{(\mu+P)^2} \hat a^{\prime} r^2  A_{\Sigma}\right]_0^{r_{\Sigma}}&=&A_{\Sigma} r_{\Sigma}^2 \hat a^{\prime}_{\Sigma} =\nonumber\\
&=&A_{\Sigma} \frac {r_{\Sigma}^2}{2} \left( \frac{T_m^{\prime}}{T_m}-\frac{P^{\prime}}{P}\right)_{r_{\Sigma}}
\end{eqnarray}
that is null because the pressure $T_m$ behaves at the boundary equal than $P$ and $P(r_{\Sigma})=0$ \cite{epjc}. Equivalently we can argue that $\hat a^{\prime}_{\Sigma}$  vanishes since the boundary condition for that interior metric function establishes that $\hat a^{\prime}_{\Sigma}=\hat \psi^{\prime}_{\Sigma}$ and it is null because the exterior metric function is the corresponding to Schwarzschild.

\vskip 2mm

\subsection{Sources of non-spherical vacuum space-time}

In this case the integrals $S_n^I$ no longer vanish (in general) because of the angular dependence of the metric functions, as well as the integrals $T_n$  from (\ref{firstfisical})-(\ref{E8pi}) incorporate those contribution missing in the spherical case. The simplest interior metric functions are \cite{weylsources} those of equation (\ref{aygsimple}) with $F=G=0$.

As a matter of ilustration let us calculate the contributions to the quadrupole moment of both kind of volume integrals over the source  for the Erez-Rosen space-time \cite{ER}
which the exterior metric function ${\displaystyle \psi=\psi^s-q_2 Q_2(x)P_2(y)}$ with arbitrary Weyl coeficient (of the Erez-Rosen representation) $q_2$, and the prolate spheroidal coordinate $x\equiv \frac rm-1$.

From equations (\ref{formulas}) we obtain the following quadrupole moment $M_2$:
\begin{eqnarray}
&&M_2=\frac{2q_2}{5\tau(\tau-2) \beta_2(\tau)}\times \nonumber\\
&\times&\left[ \int_0^{r_{\Sigma}}\frac{r^2 M^2\mu}{3(\mu+3P)} P_2(x) \left[ c_1(r)+c_2(r) \mu+c_3(r) P\right] dr +\right. \nonumber \\
&+&\left.  \int_0^{r_{\Sigma}} r^2\frac{Z}{\sqrt A} \left[2M^2P_2(x) c_4(r)+A c_5(r)  r (r-M) \right] dr \right] \nonumber\\
\label{M2ER}
\end{eqnarray}
with 
\begin{eqnarray}
c_1(r)&=&-2s \ a(\tau) +6s \ b(\tau) \nonumber \\
c_2(r)&=&-4\pi r^2\left[2(7-9s)a(\tau) +\left(9s-\frac{14}{3}\right)b(\tau)\right] \nonumber \\
c_3(r)&=&4\pi r^2\left[6(1-s)a(\tau) +(3s-2)b(\tau)\right] \nonumber \\
c_4(r)&=&(3-2s)a(\tau) +(s-1)b(\tau)  \nonumber \\
c_5(r)&=&6(1-s)a(\tau) +(3s-2)b(\tau)
\end{eqnarray}
where $s\equiv r/r_{\Sigma}$ and the notation ${\displaystyle a(\tau) \equiv  \frac{Q_2(\tau-1)}{r_{\Sigma}^2}}\quad $, ${\displaystyle b(\tau) \equiv \frac{\partial_x Q_2(x)_{(\tau-1)}}{r_{\Sigma} M}}$ has been used for the Legendre polynomial of second kind $Q_2(x)$ and its derivative, both of them evaluated in the boundary $x_{\Sigma}=\tau-1$.

\section{Conclusions}

We have shown that we are able to relate the sources of Weyl solutions  with their RMM. The procedures for the explicit calculation of the RMM of any space-time are circumscribed to the metric that describe the gravitational field. The definitions of Geroch-Hansen and Thorne involve vacuum solutions as well as the method of Fodor-Hoenselaers-Perjes (FHP), or \cite{suecos} and others developing Thorne's definition by using harmonic coordinates manage the exterior gravitational field of the compact object.

In this work a definition of RMM \cite{RMMsource} extended to whole space-time is developed explicitly for a global metric. Due to the own characteristics of that definition it was shown in \cite{RMMsource} that the RMM can be calculated by means of a flux integral at the infinity just by using the exterior metric out there. But that flux integral is equivalent to a volume integral through a generalized Gauss theorem. Nevertheless neither the interior metric  nor the source itself were used in that work since harmonic  coordinates were used to implement the flux whereas the interior metric in that system of coordinates is unknown.

Now an space-time described by a global metric is used to implement the definition from \cite{RMMsource} in such a way that relevant expressions for the RMM are obtained in terms of the material content of the source and the interior metric. Hence, those integral expressions constrained to the volume of the source allows us to calculate The RMM from the physical characteristics of the source. 

This result is providing a relevant generalization of the procedure commonly developed in newtonian gravity to define the multipole moments by means of the density of the compact object. And, at the same time a generalization of the definition of Tolman mass \cite{tolman} and Komar moments \cite{komar} is derived.

In this work it is proved that the RMM can actually be considered as physical characteristics  both of the gravitational field as well as the source conveniently matched to that exterior metric. Besides, it is the proper matching condition the guarantee to derive the volume integral expressions or the flux integrals version that lead to the relationship linking those expressions involving the source with the  RMM known in the literature associated to the gravitational field.

Consequently, from now onwards it is not necessary to know the RMM of the gravitational field matched to a  determined source but it is sufficient the energy-momentum tensor and the interior metric to calculate those RMM. Up to now the RMM gave relevant physical  information  about the source (flattening, shape, symmetry,..) and they could be connected to the orbital movement of test particles (see for instance a recent paper \cite{kopeikin} on relativistic celestial mechanics discussing the Post-Newtonian dynamics of an isolated gravitating system consisting of N extended bodies moving on a curved space-time and the relevance of multipole moments for accurate prediction of orbital dynamics of extended bodies in inspiraling binary systems
or construction of templates of gravitational waves at the merger stage when the strong gravitational
interaction between the higher-order multipoles of the bodies play a dominant role). Nevertheless, if this was the case before, and it still works so satisfactorily, from now on it may be otherwise, since the result presented here  allows us to link directly the RMM with the physics of the source.

\section*{Acknowledgments}
This  work  was  partially supported by the Spanish  Ministerio de Ciencia, Innovaci\'on y Universidades.
Subdirecci\'on General de Proyectos de Investigaci\'on, under Research Project No. PGC2018-096038-B-I00
(MINECO/FEDER), as well as  the Consejer\'\i a
de Educaci\'on of the Junta de Castilla y Le\'on under the Research Project Grupo
de Excelencia GR234 Ref.:SA096P20 (Fondos Feder y
en l\'\i nea con objetivos RIS3).


%

\end{document}